\documentclass[journal]{IEEEtran}

\usepackage{cite}
\ifCLASSINFOpdf
\else
\fi

\usepackage{graphicx}
\usepackage{amsmath,amsthm,amssymb}
\usepackage{multirow}
\usepackage{color}
\usepackage{subfig}
\usepackage{soul}
\usepackage{breqn}
\usepackage[table,xcdraw]{xcolor}
\usepackage{stackengine}
\usepackage{tabularx}
\usepackage{todonotes}
\usepackage{booktabs}
\usepackage{siunitx}
\usepackage{enumerate}

\usepackage{soul}
\usepackage{todonotes}

\begin{document}

\title{Multi-Discriminator Sobolev Defense-GAN Against Adversarial Attacks for End-to-End Speech Systems}

\author{Mohammad Esmaeilpour,~\IEEEmembership{Student Member,~IEEE,}
 Patrick Cardinal,~\IEEEmembership{Member,~IEEE,}
 and~Alessandro Lameiras Koerich,~\IEEEmembership{Member,~IEEE}
\thanks{M. Esmaeilpour, P. Cardinal, and A. L. Koerich are with the D\'{e}partement de G\'{e}nie Logiciel et des TI, \'{E}cole de Technologie Sup\'{e}rieure (\'{E}TS), Universit\'{e} du Qu\'{e}bec, Montr\'{e}al, QC, Canada,
e-mail: (mohammad.esmaeilpour.1@ens.etsmtl.ca, patrick.cardinal@etsmtl.ca, alessandro.koerich@etsmtl.ca).}
}

\markboth{Submitted to IEEE Transactions on Information Forensics and Security,~Vol.~X, No.~X, March~2021}%
{Esmaeilpour \MakeLowercase{\textit{et al.}}: IEEE Transactions on Information Forensics and Security}

\maketitle

\begin{abstract}
This paper introduces a defense approach against end-to-end adversarial attacks developed for cutting-edge speech-to-text systems. The proposed defense algorithm has four major steps. First, we represent speech signals with 2D spectrograms using the short-time Fourier transform. Second, we iteratively find a safe vector using a spectrogram subspace projection operation. This operation minimizes the chordal distance adjustment between spectrograms with an additional regularization term. Third, we synthesize a spectrogram with such a safe vector using a novel GAN architecture trained with Sobolev integral probability metric. To improve the model's performance in terms of stability and the total number of learned modes, we impose an additional constraint on the generator network. Finally, we reconstruct the signal from the synthesized spectrogram and the Griffin-Lim phase approximation technique. We evaluate the proposed defense approach against six strong white and black-box adversarial attacks benchmarked on DeepSpeech, Kaldi, and Lingvo models. Our experimental results show that our algorithm outperforms other state-of-the-art defense algorithms both in terms of accuracy and signal quality.    
\end{abstract}



%
\IEEEpeerreviewmaketitle

\section{Introduction}
\IEEEPARstart{T}{here} is a large volume of publications on applying deep learning algorithms for audio and speech classification (i.e., transcription), which report high recognition accuracy \cite{wang2020environmental,shen2019lingvo,MozillaImplementation}. During the last decade, the primary focus has been designing new architectures, for instance, variants of convolution \cite{sainath2013deep}, recurrent \cite{graves2013speech}, and attention configurations \cite{bahdanau2016end} to improve classification accuracy and model generalizability. However, it has been proven that these advanced models might undergo extreme vulnerability against carefully crafted adversarial signals both in 1D and 2D representation (spectrogram) domains \cite{carlini2018audio,esmaeilpour2019robust}.

The major focus of this paper is in response to this vulnerability issue. We have developed an adversarial defense approach against varieties of end-to-end speech-to-text attack algorithms. Toward this end, we firstly review the state-of-the-art adversarial attacks in Section~\ref{sec:backgroundAttack}. We also provide details about the background of the defense approaches in Section~\ref{sec:backgroundDefense}. Section~\ref{sec:proposed} introduces the proposed adversarial defense algorithm followed by comprehensive experimental results in Section~\ref{sec:experiements}. In summary, we make the following contributions in this paper:
\begin{enumerate}[(i)]
\item introducing a novel adversarial defense approach based on a multi-discriminator generative adversarial network (GAN) in the restricted Sobolev space~\cite{brezis2010functional};
\item establishing simple yet effective architectures for both the generator and discriminator networks;
\item developing an adjusted chordal distance with a complementary regularization term toward achieving a safe input vector for the generator model; 
\item characterizing a constraining technique for improving the stability of our generative model in adverse environmental scenarios;
\item experimentally proving the effectiveness of the proposed defense approach for white and black-box as well as targeted and non-targeted attack scenarios.
\end{enumerate}

\section{Background: Adversarial Attack}
\label{sec:backgroundAttack}
An adversarial signal $\vec{x}_{adv}$ carries inaudible perturbation $\delta$, and it can fool the victim classifier (the transcription model) toward any target phrase $\hat{\mathbf{y}}$ defined by the adversary \cite{carlini2018audio}. The actual value of $\delta$ is dependent on the length of $\hat{\mathbf{y}}$ (the number of characters) and the characteristics of the original carrier signal $\vec{x}_{org}$ ($\vec{x}_{adv}=\vec{x}_{org}+\delta$) \cite{carlini2018audio,qin2019imperceptible}. For measuring the loudness (distortion) of this perturbation relative to the carrier signal, a logarithmic-scale metric has been proposed by Carlini and Wagner~\cite{carlini2018audio}:
\begin{equation}
    l_{\text{dB}}(\vec{x}_{adv})=l_{\text{dB}}(\delta)-l_{\text{dB}}(\vec{x}_{org})
    \label{eq:log_distortion}
\end{equation}
\noindent where $l(\cdot)$ denotes the loudness of the original 1D signal $\vec{x}_{org} \in \mathbb{R}^{n\times m}$ in dB, and $n$ and $m$ denote the length and number of channels, respectively.  
For $l_{\text{dB}}(\vec{x}_{adv}) < \epsilon$ where $\epsilon$ is a small threshold, $\vec{x}_{adv}$ sounds almost seamless to $\vec{x}_{org}$ according to the 
C\&W attack for the speech-to-text model~\cite{carlini2018audio}:
\begin{equation}
   \min \left | \delta \right |_{2}^{2}+\sum_{i} c_{i}.\mathcal{L}_{i}(\vec{x}_{org,i}+\delta_{i},\pi_{i}) \quad \mathrm{s.t.} \quad l_{\text{dB}}(\vec{x}_{adv}) < \epsilon 
   \label{eq:cwa}
\end{equation}
\noindent where $c_{i}$ is a scaling coefficient for the connectionist temporal classification loss function $\mathcal{L}(\cdot)$ \cite{graves2006connectionist}. Additionally, $\pi_{i}$ denotes string tokens without duplication, which should reduce to the character alignments $\hat{\mathbf{y}}_{i}$ ($\hat{\mathbf{y}}_{i} \neq \mathbf{y}_{i}$, where the latter refers to the ground truth character alignment) \cite{carlini2018audio}. The C\&W attack has been primarily developed for the speech-to-text DeepSpeech model \cite{MozillaImplementation}, and the experiments have shown a complete collapse of this victim model against adversarial signals crafted through Eq.~\ref{eq:cwa}~\cite{carlini2018audio}.

The C\&W attack splits the input signal into 50 frames per second, and it eventually yields a universal perturbation for the entire chunks in $\vec{x}_{org}$. This operation reduces the computational complexity of the attack algorithm compared to optimizing fine-grained $\delta_{i}$ for every chunk. However, it might negatively affect the robustness of $\vec{x}_{adv}$ in a real-world environment. In other words, playing these speech chunks over the air and recording them by another microphone, involving environmental reverberating and signal echo, might easily remove the adversarial effect ($\delta$) \cite{schonherr2020imperio}. Several algorithms crafting more resilient adversarial signals in natural environments have been proposed in response to this issue. These algorithms are based on psychoacoustic loss function \cite{szurley2019perceptual}, feature vector analysis \cite{abdullah2019practical}, and employing a set of filters (band-pass, impulse response, and white Gaussian noise) \cite{yakura2018robust}. However, these approaches have been evaluated within static environments with predefined room setups, which might reduce these algorithms' generalizability in more challenging scenarios \cite{schonherr2020imperio}. 
Inspired by Athalye {\it et al.}~\cite{athalye2018synthesizing}, which introduces the expectation over transformation (EOT) to the attack optimization formulation for regularizing the cost function (similar to Eq.~\ref{eq:cwa}), many other EOT variants have been proposed for the speech domain \cite{qin2019imperceptible,schonherr2020imperio,chen2020metamorph}. These regularizations help craft more robust adversarial signals for non-static environments,
which fit in both white and black-box attack scenarios.

The EOT proposed by Qin {\it et al.}~\cite{qin2019imperceptible} is based on an acoustic room simulator, which generates artificial utterances and environmental reverberations. This algorithm is known as Robust Attack and encodes the EOT regularization into the loss function of a speech-to-text model as~\cite{qin2019imperceptible}: 
\begin{equation}
    \ell(\vec{x}_{org,i},\delta_{i},\mathbf{y}_{i})=\mathbb{E}_{t\sim \tau}\left [ \ell_{net}\left ( \mathbf{y}_{i},\hat{\mathbf{y}}_{i} \right )+\alpha \ell_{m}(\vec{x}_{org,i},\delta_{i}) \right ]
    \label{eq:adv2}
\end{equation}
\noindent where $\alpha$ is a static scaling factor, $\ell_{net} (\cdot)$ denotes the cross entropy loss and $\ell_{m} (\cdot)$ indicates the loss function for masking threshold ($\epsilon$). In fact, $\ell_{m} (\cdot)$ constrains over the normalized power spectral density function of $\vec{x}_{org}$ and contributes to the imperceptibility of the adversarial signal \cite{qin2019imperceptible}. Additionally, $\tau$ refers to the transformation set including room reverberation settings. This attack has been tested on the Lingvo speech-to-text system \cite{shen2019lingvo} and could achieve a very high fooling rate on this advanced system.

The Imperio attack proposes another variant of EOT, which implements simulated room impulse response (RIR) filters, taking advantage of a simple deep neural network (DNN) architecture \cite{schonherr2020imperio}. Additionally, this attack embeds psychoacoustic thresholding for reducing adversarial distortion similar to Qin {\it et al.}~\cite{qin2019imperceptible} (see Eq.~\ref{eq:imperio} in~\cite{schonherr2020imperio}). 
\begin{equation}
    \vec{x}_{adv}  =\arg \max_{\vec{x}_{i}}\mathbb{E}_{h\sim H_{\dim}}\left [ P(\hat{\mathbf{y}}_{i}|\vec{x}_{i,h}) \right ]
    \label{eq:imperio}
\end{equation}
\noindent where $h\in H_{\dim}$ denotes a RIR filter and $\dim$ indicates the dimension of the filter set. The Imperio is an iterative algorithm and minimizes the adversarial perturbation $\delta$ via approximating the $\nabla_{\vec{x}_{org}}=\partial \ell_{net}(\mathbf{y},\hat{\mathbf{y}})/ \partial f^{*}(\vec{x}_{org})$ where $f^{*}(\cdot)$ denotes the post activation function. In each iteration and according to the distribution of $H_{\dim}$, an adversarial candidate $\vec{x}_{adv}=\vec{x}_{i}+\kappa \nabla_{\vec{x}_{i}}$ with the learning rate $\kappa$ should satisfy $\hat{\mathbf{y}}_{i} \neq \mathbf{y}_{i}$. This procedure continues up to reach the predefined audible threshold $\epsilon$. This attack was evaluated on the Kaldi speech-to-text system \cite{povey2011kaldi}, which employs both DNN and hidden Markov model (HMM) configurations for real-time speech transcription. It has been shown that under various environmental settings, including lecture, meeting, and office rooms, the Imperio attack has considerably turned down the transcription performance of the Kaldi system~\cite{schonherr2020imperio}. 

The EOT regularization in the Metamorph adversarial attack \cite{chen2020metamorph} is similar to the RIR filtration in the Imperio algorithm with one major difference: it implements channel impulse response (CIR) to characterize potential over the air distortions on $\delta$. This attack algorithm employs $M$ pairs of microphone-speaker transmission in different distances (similar to $H_{\dim}$) to encompass a wide range of reverberations in yielding minimal perturbation:
\begin{equation}
    \arg \min_{\delta} \alpha_{t} l_{\text{dB}}(\vec{x}_{adv})+\frac{1}{M}\mathcal{L}(\vec{x}_{org}+\delta_{i},\pi_{i})
    \label{eq:metamorph}
\end{equation}
\noindent where $\alpha_{t}$ denotes a trade-off scalar between the fooling rate of the model and the signal quality. Similar to the C\&W attack, the Metamorph attack was evaluated on the DeepSpeech victim model. The experiments showed an attack success rate of around 90\% and low Mel-cepstral distortion for this white-box algorithm \cite{chen2020metamorph}.

Since integrating the EOT regularization into the adversarial optimization problem requires access to the victim model's cost function, the black-box attacks can not directly incorporate it into their formulations. For addressing this issue, a surrogate technique has been proposed and called the over-the-line approach, which provides multiple varieties of the adversarial signals to the victim model before playbacks over the air \cite{abdullah2019practical}. This operation helps the adversary to capture the environmental scene distribution without directly simulate it through reverberation filters. However, the performance of this approach is directly dependent on the comprehensiveness of the over-the-line adversarial signals. More straightforward yet effective black-box adversarial attacks, which do not incorporate EOT regularization with competitive performance on the DeepSpeech system, are the genetic algorithm attack (GAA) \cite{taori2019targeted} and multi-objective optimization attack (MOOA) \cite{khare2018adversarial}. These algorithms were tested for targeted and non-targeted attacks and achieved high fooling rates. 

While all the aforementioned adversarial attacks pose major security concerns against cutting-edge speech-to-text models, namely DeepSpeech, Kaldi, and Lingvo, there are few investigations on defense algorithms. The following section reviews the state-of-the-art defense approaches developed for counteracting white and black-box adversarial attacks.

\section{Background: Adversarial Defense}
\label{sec:backgroundDefense}
Developing defense approaches against robust adversarial attack algorithms can be very challenging due to several reasons. Firstly, standard speech signals have high dimensionality (e.g., 8 kHz), and even running effective compression techniques \cite{das2018adagio} for potentially discarding adversarial perturbations can be time-consuming in real-time speech-to-text transcription. Secondly, speech signals often have various channels for quality enhancement purposes \cite{peinado2006speech}.
Thus an adversary can optimize $\delta$ for such channel(s), which human auditory systems are less sensitive to them and more effectively fool the victim model \cite{virag1999single}. Thirdly, usually, speech signals carry environmental and microphone-speaker noises, which makes distinguishing a noisy signal from an adversarial very difficult even after band-pass filtering~\cite{hu2007comparative}.
In the following, we briefly review a couple of multiscale approaches that have been able to tackle these challenges to some extent. 

Inspired by Das {\it et al.}~\cite{das2017keeping}, a compression-based approach has been introduced for removing the potential adversarial perturbation on the speech signals \cite{das2018adagio}. This algorithm implements both adaptive multi-rate and MPEG audio layer-3 encoding for such an aim. Reported results showed the effectiveness of this approach in adverse scenarios for short-length signals \cite{das2018adagio}. Furthermore, for sophisticated adversarial signals, which have been precisely optimized through running the Robust Attack \cite{qin2019imperceptible}, this defense scheme failed to remove adversarial perturbations \cite{esmaeilpour2020class}.

Autoencoder-based defense GAN (A-GAN) \cite{latif2018adversarial} is structurally similar to the compression approach mentioned above. Instead of low-level signal filtering, it implements high-level feature transformation. The intuition behind this approach is transforming the signal into a similar recording using an autoencoder. The proposed autoencoder implements a complex architecture for reconstructing feature vectors to remove potential adversarial perturbation $\delta$. Extensive experiments of A-GAN on DeepSpeech and Lingvo systems have been reported by Esmaeilpour {\it et al.}~\cite{esmaeilpour2020class}. 

Since it has been proven that adversarial subspace is distinct from original and noisy signals \cite{esmaeilpour2020detection}, a defense GAN based on this fact has been developed by Esmaeilpour {\it et al.}~\cite{esmaeilpour2020class}. Unlike the compression approach and A-GAN approaches, this defense algorithm employs neither low nor high-level transformations for discarding adversarial perturbations directly on the signal. Instead, it uses a class-conditional GAN for computing a refined latent variable $\mathbf{z}_{i}$ for the generator network via:
\begin{equation}
    \nabla_{\mathbf{z}_{i}}\left \| \gamma\left [ G(\mathbf{z}_{i}),\mathbf{x}_{i}  \right ] \right \|_{2}^{2}
    \label{eq:chordal}
\end{equation}
\noindent where $\mathbf{z}_{i} \in \mathbb{R}^{d_{z}}$ with dimension $d_{z}$ is the random variable from $p_{z} \sim \mathcal{N}(0,0.4I)$ and $G(\cdot)$ with distribution $p_{g}$ denotes the generator network. Additionally, $\gamma[\cdot]$ is the chordal distant adjustment function between the input spectrogram $\mathbf{x}_{i}$ and $G(\mathbf{z}_{i})$. Eq.~\ref{eq:chordal} is iterative and finds the optimal latent variable $\mathbf{z}_{i}^{*}$, which not only forces $G(\mathbf{z}_{i}^{*})$ to lie in the original signal subspace, but also generate a spectrogram very similar to $\mathbf{x}_{i}$. 

The effectiveness of this class-conditional defense GAN (CC-DGAN) has been evaluated against the C\&W attack, the Robust Attack, and the GAA for both DeepSpeech and Lingvo systems \cite{esmaeilpour2020class}. However, it might fail for long-length signals (above six seconds) due to the generator network's instability in around 10k iterations. For addressing this issue, we propose two techniques: (i) introducing a multi-discriminator GAN to provide more informative gradients to the generator network; (ii) implementing such a GAN in the restricted Sobolev space \cite{brezis2010functional} and training the generator network according to the Sobolev function class with a bounded dominant measure. Since a special case of this restricted space is proportional to the 2D Fourier transform representation (spectrogram) \cite{brezis2010functional}, we can train our generative model in a much lower dimensionality compared to 1D speech signals. In the following section, we explain these steps as part of the proposed defense scheme.     

\section{Proposed Adversarial Defense Method: Sobolev Defense GAN (Sobolev-DGAN)}
\label{sec:proposed}
\begin{figure*}[htpb!]
  \centering
  \includegraphics[width=\textwidth]{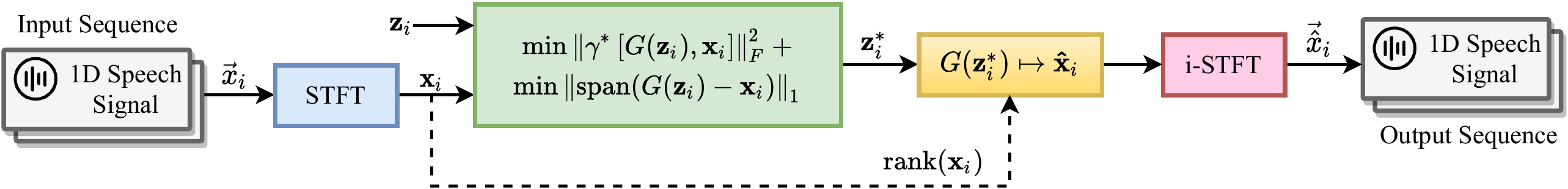}
  \caption{An overview of the proposed defense GAN approach. The 1D speech signal ($\vec{x}_{i}$) is converted to a STFT spectrogram ($\mathbf{x}_{i}$). Moreover, $\gamma\left [ \cdot \right ]$ denotes the chordal distance adjustment required for making $\mathbf{x}_{i}$ in the same subspace of the synthesized spectrogram $G(\mathbf{z}_{i})$ ($\mathbf{z}_{i} \in \mathbb{R}^{d_{z}}$ is the latent random variable). The output speech signal ($\vec{\hat{x}}_{i}$) is reconstructed using the i-STFT operation and the Griffin-Lim phase approximation approach \cite{masuyama2019deep}. Additionally, $\mathrm{rank}(\mathbf{x}_{i})$ refers to the input spectrogram's rank according to its eigenvalues computed in the Schur decomposition domain.} 
  \label{overview-defenseG}
  \vspace{-15pt}
\end{figure*}

The proposed adversarial defense approach against speech attacks has four steps, as depicted in Fig.~\ref{overview-defenseG}: (i) signal representation (conversion from 1D vector to 2D matrix) using short-time Fourier transform (STFT)~\cite{griffin1984signal}; (ii) chordal distance adjustment with a complementary regularization term for projecting the given input spectrogram onto the original subspace (the process shown in the green color); (iii) spectrogram synthesis using a Sobolev GAN and an optimal safe vector $\mathbf{z}_{i}^{*}$ (yellow block in Fig.~\ref{overview-defenseG}); (iv) inverse STFT (i-STFT) for reconstructing the speech signal.

\subsection{Spectrogram: 2D Representation of 1D Speech Signal}
\label{seq:representSpectro}
There are several standard transformations in the audio and speech processing domains for representing a signal into a 2D spectrogram, such as continuous or discrete wavelet transform, 
Mel-frequency cepstral coefficients,
and STFT. All these transformations have some advantages over each other, and they have been widely used for unsupervised, weakly supervised, and supervised learning tasks. Moreover, the highest recognition accuracies have been often reported for the models trained on these representations over 1D signals \cite{MozillaImplementation,chorowski2019unsupervised}. This is presumably due to the lower dimensionality of spectrograms and the inherent ability of these transformations in extracting more distinctive learning features compared to 1D signals \cite{deng2018speech}. 

This paper uses the STFT to generate spectrograms from the given speech signals since it is more closely related to the Sobolev integral probability metric (IPM)~\cite{mroueh2017sobolev}, which we employ to train our generator network. This metric correlates well with the Fourier coefficients encoded in the STFT spectrograms and likely helps extract more distinctive features. The theoretical approach for crafting an STFT spectrogram is as follows.

For a given discrete signal $a[n]$ with length $n$ (sampled from a 1D speech signal $\vec{x}$ in the time domain), we can define the Fourier transform using a Hann function $A[\cdot]$ as \cite{griffin1984signal}:
\begin{equation}
  \mathrm{STFT}\begin{Bmatrix} a[n] \end{Bmatrix}(k,\omega)=\sum_{n=-\infty}^{\infty}a[n]A[n-k]e^{-j\omega n}
 \label{eq:STFT}
\end{equation}
\noindent where $k$ is the shifting scale ($k\ll n$) and $\omega$ indicates the frequency coefficients. For capturing more features from $a[n]$, this operation applies on the overlapping signal chunks (i.e., 50 ms) according to a predefined sampling rate (e.g., 16 kHz). For generating the spectrogram, we need to compute the power spectrum of Eq.~\ref{eq:STFT} as:
\begin{dmath}
 \mathrm{SP_{STFT}}\begin{Bmatrix} a[n] \end{Bmatrix}(k,\omega)=
 \left | \sum_{n=-\infty}^{\infty}a[n]A[n-k]e^{-j\omega n} \right |^2
 \label{eq:STFT_spec}
\end{dmath}
\noindent where it generates a 2D matrix for a given speech signal $\vec{x}_{i}$. In the next subsection, we explain the second step of the proposed defense approach, which finds a refined $\mathbf{z}_{i}^{*}$ from the combination of a random $\mathbf{z}_{i} \in \mathbb{R}^{d_{z}}$ and the original input spectrogram ($\mathbf{x}_{i}$).

\subsection{Chordal Distance Adjustment for Spectrogram Projection}
Generally, there are two categories in developing defense approaches against adversarial attacks: running low or high-level transformations for filtering the input signal aiming at discarding potential adversarial perturbation (as discussed in Section~\ref{sec:backgroundDefense}); synthesizing a very similar signal to a given input vector without running any filtration operation \cite{esmaeilpour2020class,samangouei2018defensegan}. While most of the introduced algorithms fall into the first category, they are often less reliable since they obfuscate gradient vectors \cite{athalye2018obfuscated}. However, developing a synthesis-based defense algorithm is more challenging since it requires two key steps — a projection of the input space and a stable generative model. Since the proposed defense approach fits the second category, therefore we introduce novel techniques for these steps.

\begin{figure*}[htpb!]
  \centering
  \includegraphics[width=0.85\textwidth]{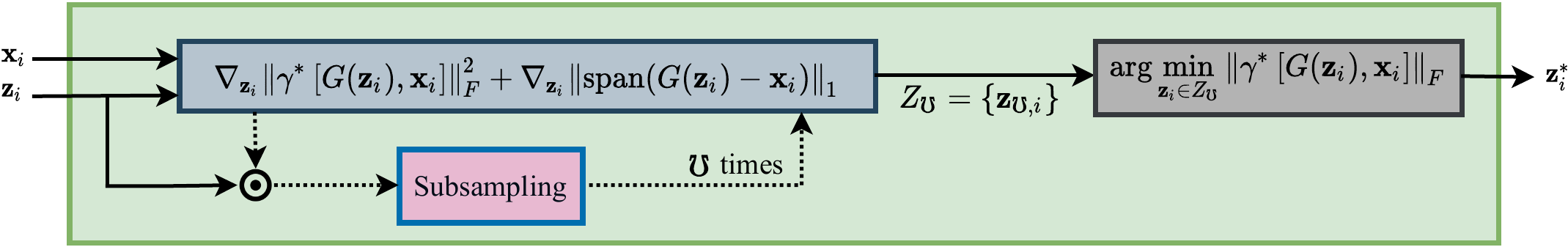}
  \caption{Overview of the proposed spectrogram subspace projection using the chordal distance adjustment and a complementary regularization term. The subsampling process is implemented with the distribution $\mathcal{N}(0.5,0.5I)$ (ratio of 0.5) for avoiding ill-conditioned pencils \cite{van1983matrix}, and a dotted line shows the internal loop. Upon producing a candidate set of $Z_{\mho}$ vectors from the given inputs, we select that $\mathbf{z}_{i}$ which minimizes the adjusted chordal distance between the synthesized spectrogram $G(\mathbf{z}_{i})$ and the input spectrogram $\mathbf{x}_{i}$.}
  \label{overview-optimizaeSubspace}
  \vspace{-15pt}
\end{figure*}

The main goal in this step is finding a safe $\mathbf{z}_{i}^{*} \in \mathbb{R}^{d_{z}}$ for the generator network according to two main conditions: $G(\mathbf{z}_{i}^{*})$ should lie in the subspace of the original signal distribution represented by $p_{r}$ (approximated by $p_{g}$); the synthesized spectrogram $G(\mathbf{z}_{i}^{*})$ should be very similar to the spectrogram of the given 1D speech signal ($\mathbf{x}_{i}$) using the $\ell_{2}$ distance metric. Toward this end, for every input spectrogram $\mathbf{x}_{i}$, we solve an optimization problem searching all possible $\mathbf{z}_{i}\in\mathbb{R}^{d_{z}}$ to find the $\mathbf{z}_{i}^{*}$ that meets the conditions above. Fig.~\ref{overview-optimizaeSubspace} shows an overview of this operation.

Inspired by Xingjun {\it et al.}~\cite{ma2018characterizing}, which proved that adversarial examples lie in distinct subspaces from original and noisy input samples, the chordal distance metric has been introduced for measuring interspaces among spectrogram manifolds \cite{esmaeilpour2020detection}. This metric, defined in the Schur decomposition domain for the triplet of original, noisy, and adversarial spectrograms, can be written as \cite{van1983matrix}:
\begin{equation}
    \mathrm{chord}(\lambda[G(\mathbf{z}_{i})], \lambda[\mathbf{x}_{i}])\leq \frac{\epsilon}{\sqrt{\begin{bmatrix}
\begin{pmatrix}\Phi^{\mathcal{H}} G(\mathbf{z}_{i})\Gamma \end{pmatrix}+\begin{pmatrix} \Phi^{\mathcal{H}} \mathbf{x}_{i}\Gamma  \end{pmatrix}\end{bmatrix}^2}}
\label{eq:chordDistance}
\end{equation}
\noindent where $\epsilon\leq 20 \mathrm{dB}$ is the maximum audible perturbation threshold, which can be defined (or optimized) by the adversary, $\lambda[\cdot]$ denotes the eigenvalue vector function class obtained with Schur decomposition. $\Gamma$, $\Phi$, and $\Phi^{\mathcal{H}}$ (conjugate transpose of $\Phi$) are random unit 2-norm operators, which satisfy \cite{van1983matrix}:
\begin{equation}
    \mathbf{x}_{i}\Gamma=\lambda[\mathbf{x}_{i}] G(\mathbf{z}_{i})\Gamma \quad \mathrm{and} \quad \Phi^{\mathcal{H}} \mathbf{x}_{i} = \lambda[G(\mathbf{z}_{i})] \Phi^{\mathcal{H}}G(\mathbf{z}_{i})
\end{equation}
\noindent For simplicity, we assume that these operators are static for all samples. Although this assumption simplifies the computation, it might result in ill-conditioned cases where an adjustment $\gamma[\cdot]$ is needed ($\mathrm{chord}(\cdot)+\gamma[\cdot]$) \cite{van1983matrix}. It has been shown that this adjustment is relatively large for adversarial spectrograms compared to original and noisy samples \cite{esmaeilpour2020detection}. Therefore, iteratively minimizing over $\gamma[\cdot]$ for $\mathrm{chord}(\lambda[G(\mathbf{z}_{i})], \lambda[\mathbf{x}_{i}])$ considerably increases the chance of finding the safe $\mathbf{z}_{i}^{*}$ that satisfies the conditions mentioned above \cite{esmaeilpour2020class,esmaeilpour2020detection}.

Since $\lambda[\cdot]$, defined in the Schur decomposition domain, is sorted (descending) and it is inductive (coefficient of both $\lambda[G(\mathbf{z}_{i})]$ and $\lambda[\mathbf{x}_{i}]$ have upper bound \cite{brezis2010functional}), according to Zorn lemma \cite{brezis2010functional} there exists a relative maximal coefficient for both $G(\mathbf{z}_{i})$ and $\mathbf{x}_{i}$ in the Hahn–Banach analytic form. Thus, we define:
\begin{equation}
    \underbrace{\gamma[\lambda[G(\mathbf{z}_{j})],\lambda[\mathbf{x}_{j}]]}_{\gamma^{*}[\cdot]} \leq \gamma[\lambda[G(\mathbf{z}_{i})],\lambda[\mathbf{x}_{i}]] \quad \mathrm{for} \quad j\ll i
    \label{eq:gammaStar}
\end{equation}
\noindent where $j$ should be chosen according to the properties of the spectrograms. However, we empirically set $j\doteq \max(i)\cdot 0.25$ to make a reasonable trade-off between spectrogram quality and computational complexity (75\% improvement). On the other hand, this operation might constitute ill-conditioned pencils (a pencil is a manifold in the closed-form of $\psi G(\mathbf{z}_{i})-\mathbf{x}_{i}$ where $\psi \propto p_{g}$ \cite{van1983matrix}) by discarding $(i-j)$ eigenvectors. To tackle this challenge, we add a complementary regularization term to the spectrogram subspace projection formulation:
\begin{equation}
    \nabla_{\mathbf{z}_{i}} \left \| \gamma^{*}\left [ G(\mathbf{z}_{i}), \mathbf{x}_{i} \right ] \right \|_{F}^{2}+\underbrace{\nabla_{\mathbf{z}_{i}}\left \| \mathrm{span}(G(\mathbf{z}_{i})- \mathbf{x}_{i}) \right  \|_{1}}_{\text{regularization}}
    \label{eq:finalOptimizationterm}
\end{equation}
\noindent where $\mathrm{span}(\cdot)$ computes a linearly independent manifold in the Schur decomposition domain from the difference between the input and synthesized spectrograms \cite{van1983matrix}. The intuition behind this regularization term is tying $G(\mathbf{z}_{i})$ as close as possible to $\mathbf{x}_{i}$ and counteracting with the potential ill-conditioned pencils imposed from $\gamma^{*}[\cdot]$. Ill-conditioned cases often happen when $\gamma^{*}[\cdot]$ is minimized, but $G(\mathbf{z}_{i})$ and $\mathbf{x}_{i}$ are not similar.

Upon solving this optimization problem (Eq.~\ref{eq:finalOptimizationterm}), we achieve a candidate set $Z_{\mho}=\left \{ \mathbf{z}_{\mho,i} \right \}$ among all the possible $\mathbf{z}_{i} \in \mathbb{R}^{d_{z}}$. Finally, we find the most optimal vector from $Z_{\mho}$ via solving for:
\begin{equation}
    \mathbf{z}_{i}^{*}:=\arg \min_{\mathbf{z}_{i} \in Z_{\mho}} \left \| \gamma^{*}\left [ G(\mathbf{z}_{i}),\mathbf{x}_{i} \right ] \right \|_{F}
    \label{eq:argzvec}
\end{equation}
\noindent where $\mathbf{z}_{i}^{*}$ is presumably refined to provide a safe input vector for the generator model. We do not directly filter the spectrograms to remove adversarial perturbation $\delta$. We find a reliable vector for a generative model to synthesize a similar spectrogram. However, the performance of all these operations is highly dependent on the generalizability and stability of the GAN model.

\subsection{Spectrogram Synthesis Using a Sobolev-GAN}
\label{subsec:specsobo}
The generative model proposed for synthesizing spectrograms is based on the vanilla GAN \cite{goodfellow2014generative} but with an integral probability metric defined in the Sobolev space \cite{brezis2010functional,mroueh2017sobolev}. Since a specific case of such a space correlates with Fourier transform, we use this measure for training our GAN on STFT spectrograms. Moreover, we introduce novel architectures for both generator and discriminator networks. For improving the generalizability and the stability of the entire model, we propose imposing a constraint on the restricted Sobolev space and incorporating multiple discriminator networks.

The task of a generator network in a GAN configuration is minimizing the discrepancies between the synthesized ($p_{g}$) and real/original ($p_{r}$) sample distributions based on a specific measure \cite{goodfellow2014generative}. The choice of such a measure is quite important since it contributes to the generalizability of the entire model (both generator and discriminator networks) \cite{arjovsky2017towards}. During the last years, many improvements have been made in designing comprehensive distance measures on top of the $\varphi$-divergence \cite{goodfellow2014generative} such as Wasserstein \cite{arjovsky2017towards}, Stein \cite{wang2016learning}, Cram\'{e}r \cite{bellemare2017cramer}, maximum mean discrepancy (MMD) \cite{dziugaite2015training,li2017mmd}, and $\mu$-Fisher IPM \cite{mroueh2017fisher}. The function which measures this discrepancy is called critic, and it can be formulated (in the closed-form) as \cite{muller1997integral}: 
\begin{equation}
    \sup_{f\in \mathcal{F}} \left [ \mathbb{E}_{G(\mathbf{x}_{i}) \sim p_{g}} f(G(\mathbf{x}_{i}))- \mathbb{E}_{\mathbf{x}_{org} \sim p_{r}} f(\mathbf{x}_{org}) \right ]
    \label{eq:discrepany}
\end{equation}
\noindent where $\mathcal{F}$ refers to the function class, which is independent of $p_{g}$ and $p_{r}$ \cite{sriperumbudur2012empirical}. For improving the GAN stability during training, restriction often applies to the critic function following the characteristics of $\mathcal{F}$ such as Lipschitz continuity ($\left \| f \right \|_{\mathrm{Lip}}\leq 1$) in Wasserstein-GAN \cite{arjovsky2017towards} and kernel Hilbert unit ball ($\left \| f \right \|_{\mathrm{Hil}}\leq 1$) in MMD-GAN \cite{li2017mmd}. Moreover, these restrictions should be inline with the properties of the training sample modality. They might result in a weak or unstable generative model, especially for sequence generation (e.g., text and speech) \cite{mroueh2017sobolev}. 

The similarity measure used for training our GAN is the Sobolev IPM, adapted for sequence-to-sequence generation \cite{mroueh2017sobolev} such as chunks of speech signals. Formally, the function class in the Sobolev space with the zero boundary condition and the dominant probability density function $\mu(\cdot)$ has the following definition \cite{brezis2010functional,mroueh2017sobolev}:
\begin{dmath}
     \mathcal{F}= \left \{ {\mathcal{X} \rightarrow \mathbb{R}^{d_{z}} \mid \mid \mathcal{X} \rightarrow L^{p_{s}}(\mathbb{R}^{d_{z}}), f \in W^{k_{s},p_{s}}(\mathcal{X},\mu),}\\
     {\mathbb{E}_{\mathbf{x}\sim \mu} \left \| \nabla_{\mathbf{x}}f(\mathbf{x}) \right \|^{2}\leq 1, \mu \sim \mathbb{P}(p_{r}, p_{g}), k_{s}, p_{s} \in \mathbb{N} } \right \} 
\end{dmath}
\noindent where $\mathcal{X} \in \mathbb{R}^{d_{z}}$ is a compact open subset, $L^{(\cdot)}$ indicates the Lebesgue norm for $1\leq p_{s}\leq\infty$, $k_{s}$ denotes the order of the critic function, and $\mathbb{P}$ is the probability density function. The special case of the function class $\mathcal{F}$ is for $p_{s}=2$ where it forms a Hilbert space $\mathcal{H}^{k_{s}}=W^{k_{s},2}$ in connection with Fourier transform as follows \cite{brezis2010functional}:
\begin{equation}
    \mathcal{H}^{k_{s}}(\cdot)\simeq \sum \alpha_{s}\left | \hat{f}(x) \right |^{2}<\infty, f\in L^{2}(\cdot)
\end{equation}
\noindent where $\alpha_{s}$ is a scalar, and $\hat{f}(x)$ refers to the Fourier series for $f(\cdot)$. Since a spectrogram is also a set of Fourier coefficients, $W^{k_{s},2}$ provides a meaningful domain for capturing local distributions of $\mathrm{SP_{STFT}}$.
We also assume $k_{s}=1$ and simplify the underlying Sobolev space as \cite{mroueh2017sobolev}:
\begin{equation}
    W^{1,2}(\mathcal{X},\mu)=\left\{f:\mathcal{X}\rightarrow\mathbb{R}^{d_{z}},\int_{\mathcal{X}}\left\|\nabla_{\mathbf{x}}f(\mathbf{x})\right\|^{2}\mu(\mathbf{x})d\mathbf{x}<\infty\right\}
    \label{eq:soboRestriction}
\end{equation}

\noindent where this restricted Sobolev space also constraints the critic function $f$ into a unit ball $\mathbb{E}_{\mathbf{x}\sim \mu} \left \| \nabla_{\mathbf{x}}f(\mathbf{x}) \right \|^{2}\leq 1$. There are numerous possible choices for defining the dominant measure $\mu(\cdot)$ according to this restricted space's properties. However, we initialize it to $\mu(\cdot)=0.5\cdot(p_{r}+p_{g})$ which is the optimal case in training a GAN \cite{mroueh2017sobolev}. Based on these explanations and using Eq.~\ref{eq:discrepany}, we can formulate the Sobolev GAN as \cite{mroueh2017sobolev}:
\begin{dmath}
    {\min_{G_{\theta_{g}}}\left [ \sup_{f_{\vartheta}, \frac{1}{N}\sum_{i=1}^{N} \left \| \nabla_{\mathbf{x}}f_{\vartheta}(\tilde{\mathbf{x}}) \right \|^{2}\leq 1} \right ]\mathcal{E}(f_{\vartheta},G_{\theta_{g}})=}\\ {\frac{1}{N}\left ( \sum_{i=1}^{N}f_{\vartheta}(G(\mathbf{z}_{i}))-\sum_{i=1}^{N}f_{\vartheta}(\mathbf{x}_{i}) \right ), \vartheta \geq 2}
    \label{eq:discreSobolov}
\end{dmath}
\noindent where the critic function $f_{\vartheta}$ follows the imposed constraint in Eq.~\ref{eq:soboRestriction}, and $\vartheta$ is the degree of the critic function. Additionally, $N$ refers to the total number of training samples, and  $\theta_{g}$ denotes the weight vectors of the generator network. Moreover, for supporting the continuity and smoothness of $f_{\vartheta}$, especially for higher-order $\vartheta$, it is recommended to define \cite{gulrajani2017improved}:
\begin{equation}
    \tilde{\mathbf{x}}_{i}=\alpha_{\vartheta} \left ( u\mathbf{x}_{r,i}+(1-u)G(\mathbf{z}_{i}) \right )
\end{equation}
\noindent where $u \sim \mathcal{U}[0,1]$ and $\alpha_{\vartheta}$ is an empirical hyperparameter (we initialize it to $\alpha_{\vartheta}=0.9$). This change of variable implicitly interpolates between $p_{r}$ and $p_{g}$ to enhance generator model stability \cite{gulrajani2017improved}. However, this enhancement is also dependent on the configurations of both the generator network in optimizing for Eq.~\ref{eq:discreSobolov} and the discriminator network, which provides gradient vectors to $G_{\theta_{g}}$.

\begin{figure*}[htpb!]
  \centering
  \includegraphics[width=0.9\textwidth]{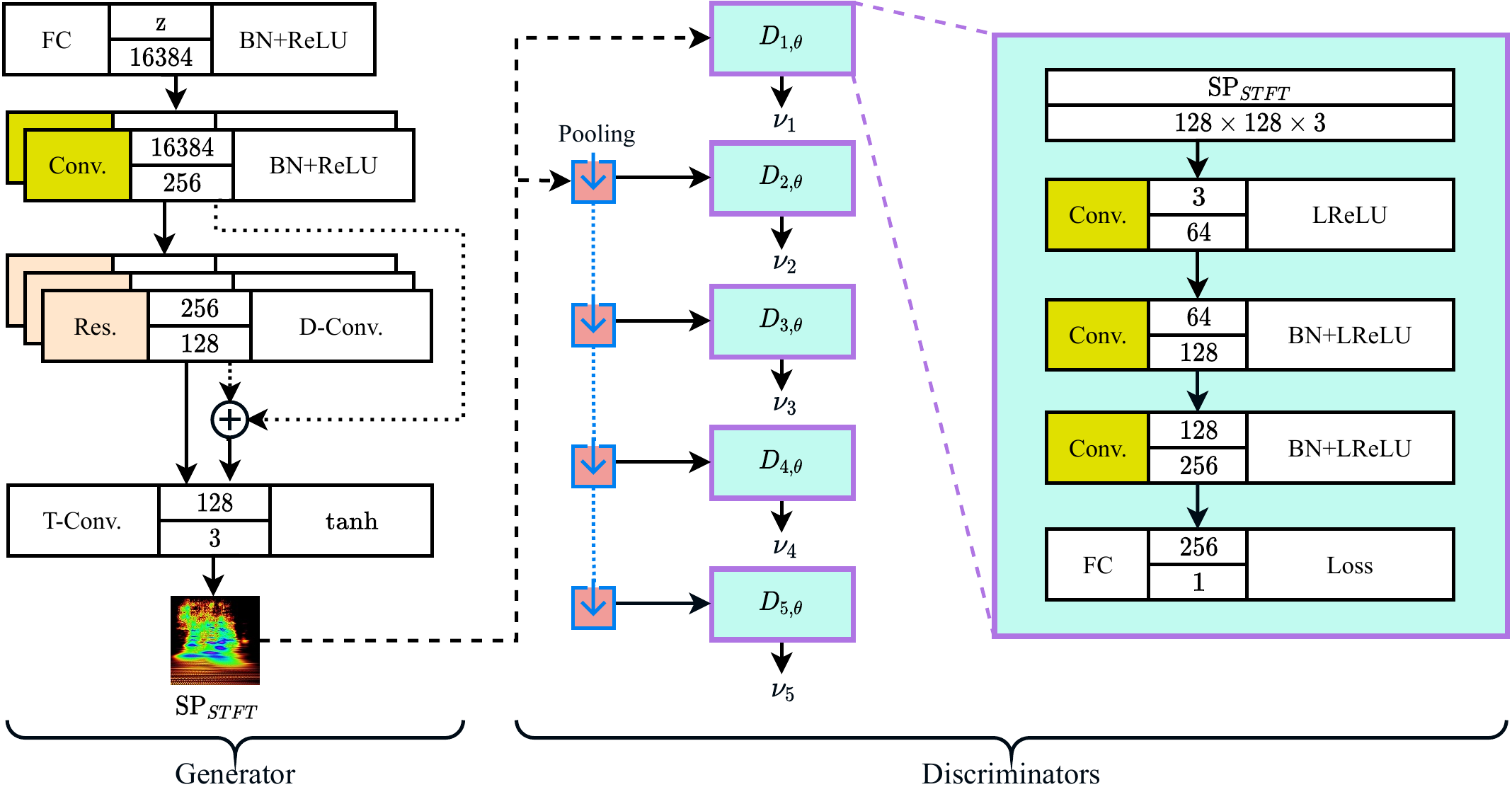}
  \caption{Overview of the proposed GAN architecture (one generator and five discriminators $D_{i,\theta}$ for $\forall i=1:5$) for spectrogram synthesis. Fully connected (FC), convolution (Conv.), dilated convolution (D-Conv.), transposed convolution (T-Conv.), and residual (Res.) layers are followed by weight normalization. The top and bottom parts of the layers refer to the input and output filters' dimensions, respectively. Moreover, $\nu_{i}$ for $\forall i=1:5$ denotes the logits of the discriminator.} 
  \label{overview-GANarchitecture}
  \vspace{-15pt}
\end{figure*}

Our proposed architecture for the generator network employs convolution and residual blocks due to their representation power in capturing continuous density functions of the input space \cite{radford2015unsupervised,brock2018large} such as spectrograms (see Fig.~\ref{overview-GANarchitecture}). The generator network contains a fully connected 1D vector layer equivalent to the total dimension of the spectrogram (128$\times$128), followed by batch normalization (BN) and rectified linear unit activation function (ReLU). This network's first hidden layers are two convolution blocks with the receptive field and stride of 5$\times$5$\times$1. The second hidden layer contains three consecutive residual blocks where each of them has dilated convolution operation with aggregation. Inspired by Kumar {\it et al.}~\cite{kumar2019melgan}, the filter size of these blocks are identical. Finally, this network's output layer is a transposed convolution \cite{mao2018effectiveness}, which yields an RGB spectrogram.

Since the discriminator network provides gradients to the generator and has a crucial role in the entire model's stability \cite{miyato2018spectral}, we empirically embedded five discriminators with identical architectures. However, we unloaded these networks from residual and long short-term memory (LSTM) blocks to avoid unnecessary complications. The filter sizes in these networks are different and, they escalate by a factor of two so that encompassing a broader range of spectrum distribution. Unlike the generator network, all the convolution layers in the discriminators deploy leaky ReLU (LReLU), as discussed in~\cite{zhang2017dilated}. The general formulation for training these GANs is:
\begin{equation}
    \min_{G} \max_{D_{i}} \mathbb{E}_{\mathbf{x}\sim p_{r}}\left [ \log D_{i}(\bold{x}) \right ]+
\mathbb{E}_{\mathbf{z}\sim p_{z}}\left [ \log \left ( 1-D_{i}(G(\bold{z})) \right ) \right ]
\end{equation}

\noindent where $\forall i=1:5$ and $p_{\mathbf{z}} \sim \mathcal{N}(0,I)$. The loss function of these networks is similar to the hinge objective function introduced in \cite{miyato2018spectral}.
However, according to the Sobolev IPM:
\begin{dmath}
    {\mathcal{L}_{S}(\vartheta,\theta_{g},\theta_{d},\varrho_{s})=\mathcal{E}(f_{\vartheta},G)+\varrho_{s}(1-\Omega_{s}(f_{\vartheta},G))}\\{-\frac{\rho}{2}(\Omega_{s}(f_{\vartheta},G)-1)^{2}}
\label{eq:lossFunc}
\end{dmath}
\noindent and in this definition, $\Omega_{s}(\cdot)$ in the restricted Sobolev space $W^{1,2}(\mathcal{X},\mu)$ is differentiable and regarding Eq.~\ref{eq:discreSobolov}, it is defined as \cite{mroueh2017sobolev}: 
\begin{dmath}
    \frac{1}{2N}\left ( \sum_{i=1}^{N}\left \| \nabla_{\mathbf{x}}f_{\vartheta}(\mathbf{x}_{i}) \right \|^{2}+ \sum_{i=1}^{N}\left \| \nabla_{\mathbf{x}}f_{\vartheta}(G(\mathbf{z}_{i})) \right \|^{2} \right )
\end{dmath}
\noindent Moreover, $\varrho_{s}$, $\theta_{d}$, and $\rho >0$ denote the Lagrange multiplier, the weight vectors of each discriminator network, and the penalty weight for providing higher smoothness in training, respectively \cite{mroueh2017fisher}. One potential side effect of training the generator with multiple discriminators is the difficulty of making a trade-off between sample variety an quality. For tackling this challenge, we use orthogonal regularization (OR) for all the discriminator networks using a simple linear similarity measure \cite{brock2017neural}:
\begin{equation} 
R_{\varpi}=\varpi\left \| \theta_{d}^{\top} \theta_{d}\odot (\bold{1}-I)\right \|^{2}_{F} \label{eq:orthoReg}
\end{equation}
\noindent where empirically $\varpi\in(10^{-5},10^{-5}]$ is a small tuning coefficient, and $\bold{1}$ indicates a matrix with constant values of one \cite{brock2018large}. This regularization forces the discriminator network to reduce dissimilarity among filters to learn more distinctive features. However, this might negatively affect the generator performance in capturing all the possible modes from the spectrogram, cause instability in a higher number of iterations, and generate oversmoothed samples \cite{esmaeilpour2020unsupervised}. In response to this issue, we propose a new constraint for the critic function $f_{\vartheta}$ as the following.\\

\noindent \textbf{\textit{Proposition:}} There is an achievable upperbound (supremum) for the continuous (and partially differentiable) critic function $f_{\vartheta}(\cdot)$ in the restricted Sobolev space $W^{1,2}(\mathcal{X},\mu)$ with:
\begin{equation}
    L^{\eta}(\mathcal{X})=\left \{ f_{\vartheta}: \mathcal{X}\rightarrow \mathbb{R}, \left | f_{\vartheta} \right |^{\eta}\in L^{1}(\mathcal{X}) \right \}
\end{equation}
\noindent where $\left \| f \right \|_{L^{1}}=\left \| f \right \|_{1}$ and $1 \leq \eta \leq \infty$. This reduces the space definition in Eq.~\ref{eq:soboRestriction} to $\int_{\mathcal{X}}\left \| \nabla_{\mathbf{x}}f(\mathbf{x}) \right \|^{2}\mu(\mathbf{x})d\mathbf{x} \leq c_{\Upsilon}$ where $c_{\Upsilon}$ is a positive static scalar.\\

\noindent \textbf{\textit{Proof:}} According to the rigid constraint $\mathbb{E}_{\mathbf{x}\sim \mu} \left \| \nabla_{\mathbf{x}}f(\mathbf{x}) \right \|^{2}\leq 1$ imposed on $W^{1,2}(\mathcal{X},\mu)$ in Eq.~\ref{eq:soboRestriction}, it always supports $\left \| \nabla_{\mathbf{x}}f(\mathbf{x}) \right \|^{2} \in L^{\eta}$ (the Lebesgue norm). If we bind $\mu(\mathbf{x}) \in L^{{\eta}'}$ where $\eta{}'$ denotes the conjugate exponent of $\eta$ ($1/\eta+1/\eta{}'=1$), then using the H{\"o}lder's inequality \cite{brezis2010functional}, we can write:
\begin{equation}
    \int_{\mathcal{X}}\left \| \nabla_{\mathbf{x}}f(\mathbf{x}) \right \|^{2}\mu(\mathbf{x})d\mathbf{x}\leq \underbrace{\left \| \nabla_{\mathbf{x}}f(\mathbf{x}) \right \|_{\eta}^{2}\left \| \mu(\mathbf{x}) \right \|_{\eta{}'}}_{c_{\Upsilon}< < \infty}\quad\square
    \label{eq:cgammConst}
\end{equation}
\noindent where $c_{\Upsilon}$ is dependent on the cumulative distribution of $\mu(\mathbf{x})$. This constraint forces the generator network to discard local sample distributions which lie far from the optimal generator distribution ($\left \lceil p_{r}+p_{g} \right \rceil/2$). It also implicitly helps the discriminator network avoid shattering gradients vectors since the learning space bound to $c_{\Upsilon}$. 

For synthesizing a spectrogram similar to the given $\mathbf{x}_{i}$, the generator network maps the safe vector $\mathbf{z}_{i}^{*}$ onto $\hat{\mathbf{x}}_{i}$ and then tunes the generated spectrogram with the $\mathbf{x}_{i}$'s rank \cite{van1983matrix} in the Schur decomposition domain. Even if this tuning is optional, it improves the quality of $\hat{\mathbf{x}}_{i}$ and reduces the potential dissimilarity between $G(\mathbf{z}_{i}^{*})$ and $\mathbf{x}_{i}$.

The last step of the proposed adversarial defense approach is transforming the synthesized spectrograms into the time domain using the inverse STFT operation. This step is necessary only for end-to-end speech-to-text victim models upon adversary's discern.

Reconstructing an audio or speech signal from a spectrogram requires the associated phase vectors from the transformation function (e.g., STFT). There are two main approaches for such an aim: using original phase vectors and approximating phase vectors. Obviously, in the first approach, the reconstructed signals' quality will be very similar to the original counterparts since they share the same timing. However, original phase vectors might not always be accessible, contrary to the second signal reconstruction approach. On the other hand, approximated phase vectors usually add audible noise to the reconstructed signal and degrade its quality. Therefore we opted for the second approach since accessing the original phase vectors might be prohibitive in some senses. Specifically, we use the recognized Griffin-Lim algorithm for the i-STFT procedure~\cite{masuyama2019deep}. Since this may raise concerns about the quality of the reconstructed signals, we measure their peculiarity with some metrics. 

\section{Experimental Results}
\label{sec:experiements}
In this section, we analyze the proposed defense scheme's performance from two points of view: the defense algorithm's success rate by measuring the word error rate and sentence-level accuracy scores, and the quality of the signals from the synthesized spectrograms and the approximated phase vectors. The latter also includes comparing signals after filtration by various defense algorithms. This shows the impact of defense algorithms on speech signals.

Our benchmarking victim models are DeepSpeech, Kaldi, and Lingvo, which employ both the conventional and cutting-edge learning blocks, such as HMM, convolutional, recurrent, LSTM, and residual configurations. These models are trained on Mozilla common voice (MCV) \cite{MozillaCommonVoiceDataset} and LibriSpeech \cite{panayotov2015librispeech} comprehensive datasets, including numerous utterances. Moreover, they contain above 1,000 hours of recordings organized in short ($\leq 6$ sec) and long ($>6$ sec) voice clips.

In all our experiments, we use a combination of strong white and black-box end-to-end adversarial attacks, as discussed in Section~\ref{sec:backgroundAttack}. For every adversarial signal, regardless of EOT type, we assign ten targeted incorrect different phrases, including silence \cite{carlini2018audio}, and five non-targeted incorrect random phrases with different lengths to more effectively challenge defense approaches. Meanwhile, we take identical assumptions for those algorithms that require environmental settings such as CIR and RIR filter sets for fairness in comparison. Following a common practice in adversarial studies \cite{carlini2018audio, qin2019imperceptible, taori2019targeted}, we also craft adversarial signals for a group of randomly selected portions (with shuffling) of the datasets mentioned above. More specifically, we randomly choose 25k English-speaking samples from both MCV and LibriSpeech with an almost equal number of genders (male and female), accent (e.g., United States, England, etc.), and age (the majority between 19 to 39 regarding the dataset limitation). We assign almost 60\% of these samples for training, tuning, and validating our generative model. Hence, the remaining portion will be used for developing adversarial signals using six attack algorithms.

Since we train our GAN on the spectrograms, we firstly convert speech signals into $\mathrm{SP_{STFT}}$
with a sampling rate of 22.05 kHz. Additionally, we set the total number of Mel-frequency coefficients to 20 per frame with an overlapping ratio of 0.5 and the hop length of 512. The Hann window length is initialized to 2048 with reflect padding.

We discard checkpoints with unstable learning curves during training and opt to early stop when any signs of instability become present \cite{brock2018large}. For all the architectures (the generator and five discriminators) we use Adam optimizer with a static learning rate of $2\cdot 10^{-5}$ and hyperparameters $\beta_{1}=0$ and $\beta_{2}=0.9$. We empirically set the required number of steps for the generator network over the discriminators to two with a decay ratio of $0.99$ on four NVIDIA GTX-1080-Ti and two 64-bit Intel Core-i7-7700 (3.6 GHz) with 8$\times$11GB and 2$\times$64GB memory, respectively. 

For evaluating the performance of the proposed defense algorithm against adversarial attacks, we also use the word error rate (WER) and sentence level accuracy (SLA)~\cite{qin2019imperceptible}. The first metric measures the summation of total phrase insertion, substitution, and deletion over the ground-truth phrases ($\mathbf{y}_{i}$). The second metric measures the ratio of correctly transcripted phrases over the total number of test speech signals. To avoid bias in our analysis, we repeat each experiment 10 times and report the average WER and SLA for each defense algorithm. Table~\ref{table:comparisonSob} summarizes the achieved results.

\begin{table*}[ht]
\centering
\caption{Comparison of the defense algorithms against strong white and black-box adversarial attacks for the DeepSpeech, Kaldi, and Lingvo victim speech-to-text models. Whereas WER and LLR, higher values for the SLA, PESQ, segSNR, and STOI metrics are better. The difference between Sobolev-DGAN$^{*}$ and Sobolev-DGAN is the latter does not incorporate the constraint proposition (Eq.~\ref{eq:cgammConst}) mentioned in Section \ref{subsec:specsobo}. Outperforming results are shown in boldface.}
\begin{tabular}{c||c||r||c|c|c|c|c|c|c}
\hline
Model                        & Attack                         & \multicolumn{1}{c}{Defense}            & Average $\mho$ & WER (\%)             & SLA (\%)            & PESQ   & segSNR  & STOI   & LLR    \\ \hline \hline
\multirow{20}{*}{DeepSpeech} & \multirow{5}{*}{C\&W}          & Compression  \cite{das2018adagio}      & $-$          & $19.14\pm 2.36$   & $49.26 \pm 2.67$ & $1.64$ & $09.31$  & $0.85$ & $0.44$ \\ \cline{3-10} 
                             &                                & A-GAN              & $-$          & $26.32 \pm 3.03$  & $36.21 \pm 0.12$ & $1.15$ & $06.95$ & $0.87$ & $0.41$ \\ \cline{3-10} 
                             &                                & CC-DGAN            & $-$          & $14.52 \pm 1.16$  & $61.23 \pm 1.02$ & $2.01$ & $12.56$ & $0.89$ & $0.38$ \\ \cline{3-10} 
                             &                                & \bf Sobolev-DGAN       & $163$          & $07.61\pm 0.47$   & $76.15 \pm 2.18$ & $2.36$ & $18.73$ & $0.91$ & $0.31$ \\ \cline{3-10} 
                             &                                & \bf  Sobolev-DGAN$^{*}$ & $159$          & $\bold{04.21 \pm 1.39}$  & $\bold{79.24 \pm 1.17}$ & $\bold{2.71}$ & $\bold{19.91}$ & $\bold{0.95}$ & $\bold{0.30}$ \\ \cline{2-10} 
                             & \multirow{5}{*}{Metamorph}     & Compression  \cite{das2018adagio}      & $-$          & $21.54\pm 2.17$   & $51.57 \pm 1.91$ & $1.55$ & $10.34$ & $0.76$ & $0.48$ \\ \cline{3-10} 
                             &                                & A-GAN              & $-$          & $19.81 \pm 3.72$  & $58.39 \pm 0.49$ & $1.59$ & $10.86$ & $0.83$ & $0.32$ \\ \cline{3-10} 
                             &                                & CC-DGAN            & $-$          & $11.89 \pm 1.23$  & $71.94 \pm 1.56$ & $1.96$ & $11.08$ & $0.85$ & $0.35$ \\ \cline{3-10} 
                             &                                &  \bf Sobolev-DGAN       & $039$          & $09.37 \pm 1.12$  & $75.19 \pm 2.18$ & $2.17$ & $14.76$ & $0.88$ & $0.34$ \\ \cline{3-10} 
                             &                                &  \bf Sobolev-DGAN$^{*}$ & $027$          & $\bold{06.79 \pm 0.19}$  & $\bold{80.34 \pm 3.67}$ & $\bold{2.45}$ & $\bold{16.01}$ & $\bold{0.93}$ & $\bold{0.31}$ \\ \cline{2-10} 
                             & \multirow{5}{*}{GAA}           & Compression  \cite{das2018adagio}      & $-$          & $27.41 \pm 3.61$  & $43.71\pm 1.32$  & $2.14$ & $14.37$ & $0.87$ & $0.39$ \\ \cline{3-10} 
                             &                                & A-GAN              & $-$          & $29.49 \pm 5.26$  & $40.88 \pm 5.37$ & $1.66$ & $12.53$ & $0.88$ & $0.37$ \\ \cline{3-10} 
                             &                                & CC-DGAN            & $-$          & $14.98 \pm 3.56$  & $69.46 \pm 2.37$ & $2.03$ & $13.52$ & $0.90$ & $0.34$ \\ \cline{3-10} 
                             &                                &  \bf Sobolev-DGAN       & $101$          & $09.68 \pm 2.73$  & $\bold{73.98 \pm 0.77}$ & $\bold{2.39}$ & $16.02$ & $0.93$ & $\bold{0.29}$ \\ \cline{3-10} 
                             &                                &  \bf Sobolev-DGAN$^{*}$ & $097$          & $\bold{05.01 \pm 0.11}$  & $72.88 \pm 4.28$ & $2.38$ & $\bold{18.91}$ & $\bold{0.94}$ & $0.30$ \\ \cline{2-10} 
                             & \multirow{5}{*}{MOOA}          & Compression  \cite{das2018adagio}      & $-$          & $17.06 \pm 0.19$  & $55.16 \pm 3.86$ & $1.87$ & $\bold{19.42}$ & $\bold{0.92}$ & $\bold{0.38}$ \\ \cline{3-10} 
                             &                                & A-GAN              & $-$          & $18.74 \pm 43.21$ & $53.07 \pm 3.06$ & $1.85$ & $14.63$ & $0.87$ & $0.41$ \\ \cline{3-10} 
                             &                                & CC-DGAN            & $-$          & $15.69 \pm 1.97$  & $61.11 \pm 2.99$ & $1.99$ & $17.81$ & $0.89$ & $0.39$ \\ \cline{3-10} 
                             &                                &  \bf Sobolev-DGAN       & $051$          & $12.25 \pm 2.84$  & $68.84 \pm 1.56$ & $\bold{2.46}$ & $19.35$ & $0.90$ & $0.36$ \\ \cline{3-10} 
                             &                                &  \bf Sobolev-DGAN$^{*}$ & $049$          & $\bold{04.23 \pm 2.32}$   & $\bold{79.36 \pm 2.16}$ & $2.30$ & $18.06$ & $0.91$ & $0.39$ \\ \hline \hline
\multirow{5}{*}{Kaldi}       & \multirow{5}{*}{Imperio}       & Compression   \cite{das2018adagio}     & $-$          & $16.29 \pm 5.17$  & $56.42 \pm 6.11$ & $\bold{2.42}$ & $15.79$ & $0.83$ & $\bold{0.32}$ \\ \cline{3-10} 
                             &                                & A-GAN              & $-$          & $17.76 \pm 0.16$  & $54.28 \pm 1.90$ & $1.23$ & $09.76$ & $0.74$ & $0.48$ \\ \cline{3-10} 
                             &                                & CC-DGAN            & $-$          & $10.19 \pm 2.93$  & $ 69.62\pm 2.63$ & $1.84$ & $\bold{16.53}$ & $0.78$ & $0.45$ \\ \cline{3-10} 
                             &                                &  \bf Sobolev-DGAN       & $093$          & $06.78 \pm 0.91$   & $75.33 \pm 2.97$ & $1.96$ & $13.98$ & $0.81$ & $0.41$ \\ \cline{3-10} 
                             &                                &  \bf Sobolev-DGAN$^{*}$ & $047$          & $\bold{03.29 \pm 1.14}$  & $\bold{82.37 \pm 3.62}$ & $2.35$ & $16.52$ & $\bold{0.89}$ & $0.35$ \\ \hline \hline
\multirow{5}{*}{Lingvo}      & \multirow{5}{*}{Robust Attack} & Compression  \cite{das2018adagio}      & $-$          & $21.56 \pm 4.15$  & $55.11 \pm 3.05$ & $2.06$ & $15.08$ & $0.74$ & $0.33$ \\ \cline{3-10} 
                             &                                & A-GAN              & $-$          & $17.90 \pm 4.21$  & $59.98 \pm 1.38$ & $\bold{2.17}$ & $14.43$ & $0.72$ & $0.34$ \\ \cline{3-10} 
                             &                                & CC-DGAN            & $-$          & $14.46 \pm 0.35$  & $64.16 \pm 2.14$ & $1.71$ & $11.09$ & $0.79$ & $0.28$ \\ \cline{3-10} 
                             &                                &  \bf Sobolev-DGAN       & $114$          & $11.99 \pm 2.76$  & $69.33 \pm 0.81$ & $1.92$ & $12.25$ & $0.76$ & $0.34$ \\ \cline{3-10} 
                             &                                &  \bf Sobolev-DGAN$^{*}$ & $136$          & $\bold{05.86 \pm 1.64}$  & $\bold{83.46 \pm 2.27}$ & $1.96$ & $\bold{17.07}$ & $\bold{0.81}$ & $\bold{0.22}$ \\ \hline
\end{tabular}
\label{table:comparisonSob}
\vspace{-15pt}
\end{table*}

Table~\ref{table:comparisonSob} shows that for most cases, the proposed defense approach (Sobolev-DGAN$^{*}$) and its variant without employing the constraining proposition (Sobolev-DGAN) introduced in Section~\ref{subsec:specsobo} outperform other defense algorithms against six strong end-to-end speech attacks. Averaged over all the conducted experiments on the three victim speech-to-text models, Sobolev-DGANs have similar performance on white (C\&W, Metamorph, Imperio, and Robust Attack) and black-box (GAA and MOOA) attack algorithms. That indicates the independence of our defense algorithm to the adversarial attack scenarios. Moreover, the total number of required iterations ($\mho$) toward achieving the safe input vector $\mathbf{z}_{i}^{*}$ for the C\&W attack and the Robust Attacks is relatively more than others. That could be interpreted as the higher power of these attacks in yielding more destructive adversarial signals since they demand an additional cost for our defense algorithm to find the input vector. However, any discussion on the resiliency of adversarial attacks and their potentials in optimizing upscale examples is beyond this paper's scope.

Furthermore, Table~\ref{table:comparisonSob} also proves the effectiveness of the proposed constraining technique for the critic function $f_{\vartheta}$ as discussed in  Section~\ref{subsec:specsobo}. Except for the GAA, Sobolev-DGAN$^{*}$ has shown higher SLA than the Sobolev-DGAN on all the victim speech-to-text models.  

For evaluating the potential negative impact of running defense algorithms on the crafted adversarial signals, we use four objective speech quality metrics: perceptual evaluation of speech quality (PESQ) \cite{rix2001perceptual}, segmental signal to noise ratio (segSNR) \cite{baby2019sergan}, short-term objective intelligibility (STOI) \cite{taal2011algorithm}, and log-likelihood ratio (LLR) \cite{baby2019sergan}. The first metric is based on cognitive modeling, and the input filter set aligns with identifying noisy intervals (high-level quality analysis). The second metric is the enhanced version of the conventional signal-to-noise ratio in audible logarithmic scale for chunks of speech signals (low-level quality analysis). The third metric evaluates the ratio of band-pass local noise perceptibility to the entire signal chunks. Unfortunately, these metrics are not normalized in a scaled interval. However, there is a direct relationship between their magnitudes and signal quality. The fourth metric is associated with a logarithmic noise ratio relative to the ground-truth scaled between $[0,1]$. Therefore, high-quality signals have lower LLR. As shown in Table~\ref{table:comparisonSob}, for the most cases, averaged over ten times experiment repetitions, both the Sobolev-DGAN$^{*}$ and Sobolev-DGAN outperform others in keeping the quality of the signals after running the defense filtration.

\begin{figure}[htpb!]
  \centering
  \includegraphics[width=0.45\textwidth]{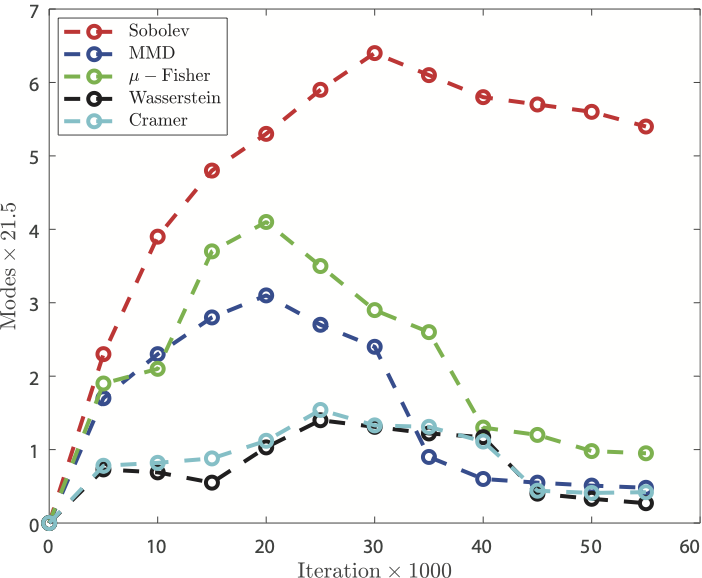}
  \caption{Monitoring the average learned modes (per batch size of 2$\times$512) by our GAN model during training on $\mathrm{SP_{STFT}}$ with different IPMs indicates potential collapse over the total number of iterations.}
  \label{fig:IPMcomp}
\end{figure}

In Section~\ref{subsec:specsobo}, we mentioned that $W^{k_{s},2}$ provides a meaningful (and comprehensive) domain for capturing local distributions of spectrograms. To investigate this claim, Fig.~\ref{fig:IPMcomp} shows the relation between the Sobolev IPM and extracted local and global probability distributions from spectrograms compared to others. Toward this end, inspired by Mao {\it et al.}~\cite{mao2018effectiveness}, we compare the mode collapse issue between the GANs trained with various IPMs as mentioned in Section~\ref{subsec:specsobo}. We have used an identical architecture for all generative models (generator and discriminators depicted in Fig.~\ref{overview-GANarchitecture}) for fairness in comparison. Additionally, we have used the same settings for these networks.

Fig.~\ref{fig:IPMcomp} shows that the average number of learned modes has an increasing behavior of up to 20k iterations for MMD and $\mu$-Fisher IPMs. For Wasserstein and Cram\'{e}r IPMs, this behavior reaches around 26k iterations. Among these, the Sobolev IPM keeps its incremental behavior up to 30k iteration with considerable bias (along the $y$-axis). That demonstrates the higher performance of $f_{\vartheta}$ in capturing the local distribution of spectrograms in the restricted Sobolev space compared to other IPM. However, it does not immune our generative model against the mode collapse issue. As depicted in Fig.~\ref{fig:IPMcomp}, our GAN gradually starts losing sample modes after 31k iterations. For tackling this issue, we used OR, spectral normalization \cite{miyato2018spectral}, and early stopped at checkpoints before the collapse.

Since there is a direct relationship between stability and generalizability of the GAN and our proposed defense algorithm, even a partially unstable generator network might result in absolute divergence in the chordal distance adjustment operation. In other words, if the GAN model is not comprehensive enough in terms of the number of learned modes, the process shown in Fig.~\ref{overview-optimizaeSubspace} might never converge. This poses more concerns for long signals with too much environmental noise. Additionally, for multi-speaker speech signals, our proposed Sobolev-DGANs not only might not be able to learn enough modes but also might recover adversarial perturbation after the i-STFT procedure. We believe that employing more constraining conditions on both the generator and discriminators may improve model stability. Moreover, conditioning the discriminator networks aligned with time-distributed filter sets can provide more distinctive features for the discriminator network to resolve the multi-speaker issue. We are determined to address these issues in future work.

\section{Conclusion}
\label{sec:conclusion}
In this paper, we proposed a novel approach for defensing speech-to-text models against end-to-end adversarial attacks. Our approach is based on reconstructing signals from synthesized spectrograms and approximated phase vectors. For spectrogram synthesis, we use a multi-discriminator GAN defined in the restricted Sobolev space. Our GAN generator network requires a safe input vector achievable through an iterative spectrogram subspace projection operation using the adjusted chordal distance. To improve our implemented generative model's performance, we impose a constraint for the critic function that learns discrepancies between real and synthesized sample distributions. We evaluated our defense approach against six strong white and black-box adversarial attacks on advanced DeepSpeech, Kaldi, and Lingvo victim models. The proposed defense approach, averaged over the total number of experiments, outperformed other algorithms according to WER and SLA metrics. Furthermore, we used four objective quality metrics for measuring the impact of running defense algorithms on speech signals. For the majority of the cases, our defense approach demonstrated higher signal quality compared to other algorithms. 

\section*{Acknowledgment}
This work was funded by the Natural Sciences and Engineering Research Council of Canada (NSERC) under Grant RGPIN 2016-04855 and Grant RGPIN 2016-06628.

\bibliographystyle{IEEEtran}
\bibliography{IEEEabrv,mybib}

\vfill


\end{document}